    \documentclass[preprint2]{aastex}

\usepackage[svgnames]{xcolor}
\usepackage[colorlinks=true,linkcolor=blue, citecolor=blue]{hyperref}

\shorttitle{A new contact binary?}
\shortauthors{Thirouin A., Sheppard S.S., and Keith S. Noll}

\begin{document}

\title{2004~TT$_{357}$: A potential contact binary in the Trans-Neptunian belt}

\author{Audrey Thirouin\altaffilmark{1}}
\affil{Lowell Observatory, 1400 W Mars Hill Rd, Flagstaff, Arizona, 86001, USA. }
\email{thirouin@lowell.edu}
 
\author{Scott S. Sheppard\altaffilmark{2}}
\affil{Department of Terrestrial Magnetism (DTM), Carnegie Institution for Science, 5241 Broad Branch Rd. NW, Washington, District of Columbia, 20015, USA.}
 \email{ssheppard@carnegiescience.edu}

 \and

\author{Keith S. Noll\altaffilmark{3}}
\affil{NASA Goddard Space Flight Center (NASA-GSFC), 8800 Greenbelt Rd, Greenbelt, Maryland, 20771, USA.}
 \email{keith.s.noll@nasa.gov}
 
\begin{abstract} 

We report photometric observations of the trans-Neptunian object 2004~TT$_{357}$ obtained in 2015 and 2017 using the 4.3~m Lowell's Discovery Channel Telescope. We derive a rotational period of 7.79$\pm$0.01~h and a peak-to-peak lightcurve amplitude of 0.76$\pm$0.03~mag. 2004~TT$_{357}$ displays a large variability that can be explained by a very elongated single object or can be due to a contact/close binary. The most likely scenario is that 2004~TT$_{357}$ is a contact binary. If it is in hydrostatic equilibrium, we find that the lightcurve can be explained by a system with a mass ratio q$_{min}$=0.45$\pm$0.05, and a density of $\rho_{min}$=2~g~cm$^{-3}$, or less likely a system with q$_{max}$=0.8$\pm$0.05, and $\rho_{max}$=5~g~cm$^{-3}$. Considering a single triaxial ellipsoid in hydrostatic equilibrium, we derive a lower limit to the density of 0.78~g~cm$^{-3}$, and an elongation (a/b) of 2.01 assuming an equatorial view. From \textit{Hubble Space Telescope} data, we report no resolved companion orbiting 2004~TT$_{357}$. Despite an expected high fraction of contact binaries in the trans-Neptunian belt, 2001~QG$_{298}$ is the unique confirmed contact binary in the trans-Neptunian belt, and 2004~TT$_{357}$ is only the second candidate to this class of systems, with 2003~SQ$_{317}$. 

\end{abstract}


\section{Introduction}

Separated Trans-neptunian binaries have a large variety of properties, from tiny satellites around large primaries to nearly equal-sized systems with primaries and secondaries having comparable sizes, and from short to long orbital periods \citep{Noll2008}. The discovery of binary systems in the trans-Neptunian belt is subject to observational limitations. Only widely separated and nearly equal-sized binaries are detected from the ground \citep{Sheppardetal2012, Parker2011, Noll2008, Veillet2002}. 

The most prolific tool for detecting binary or multiple systems is the \textit{Hubble Space Telescope (HST)} \citep{Noll2008}. However, contact binaries, and binaries with tight orbits are impossible to resolve and identify with \textit{HST} because of the small separation between the system's components. Only detailed lightcurves with a characteristic V-/U-shape at the minimum/maximum of brightness and a large amplitude can identify close or contact binaries. 

In October 2004, using the 4~m Mayall telescope (Kitt Peak, Arizona, USA), \citet{BuieMPC2004} discovered 2004~TT$_{357}$. With a semi-major axis\footnote{Semi-major axis, eccentricity and inclination are from the Minor Planet Center (MPC, February 2017).} of 55.17~AU, an inclination of 8.99$^\circ$, and an eccentricity of 0.43, 2004~TT$_{357}$ is a trans-Neptunian Object (TNO) trapped in the 5:2 mean motion resonance with Neptune\footnote{The dynamical classification is based on the Deep Ecliptic Survey Object Classifications: \url{http://www.boulder.swri.edu/~buie/kbo/astrom/04TT357.html}}. 

2004~TT$_{357}$ is a moderately red object with a slope S=14.3$\pm$2, and its optical colors are: g'-r'= 0.74$\pm$0.03~mag, r'-i'= 0.27$\pm$0.04~mag, and g'-i'=0.99$\pm$0.04~mag \citep{Sheppard2012}. With an absolute magnitude of H$_{r'}$=7.42$\pm$0.07~mag, the estimated diameter of 2004~TT$_{357}$ is 218$\pm$7~km (87$\pm$3~km) assuming an albedo of 0.04 (0.25) \citep{Sheppard2012}.

We report photometric observations of 2004~TT$_{357}$ obtained in 2015 and 2017 using the 4.3~m Lowell's Discovery Channel Telescope. The lightcurve of 2004~TT$_{357}$ presents an extreme variability that can be explained by a very elongated single object or by a contact/close binary. This paper is divided into six sections. In the next section, we present the fraction of contact binaries among the Solar System. Section 3 describes the observations and the data set analyzed. In Sections 4 and 5, we present and discuss our main results. Section 6 will present the search for companions around 2004~TT$_{357}$ with the \textit{Hubble Space Telescope}. Finally, Section 7 is dedicated to the summary and the conclusions of this work.  


\section{Fraction of contact binaries}

The extended definition of a contact/close binary system is an object consisting of two lobes in contact (bi-lobed object with a peanut/bone shape), and system of two separated objects almost in or in contact. This kind of peculiar system/object is found across the small body populations, from the Near-Earth Objects population to the trans-Neptunian belt \citep{Benner2015, Mann2007, Sheppard2004}. The expected fraction of contact binaries is high in all of the small body populations, with estimates up to 20$\%$. In the case of the Trojans larger than $\sim$12~km, the estimate is 14$\%$-23$\%$, and 30$\%$-51$\%$ for Hildas larger than about 4~km based on preliminary estimates from \citet{Sonnett2015}. Finally, \citet{Ryan2017} found that 6-36$\%$ of the Trojans are contact binaries. Several studies suggest that between 10\% and 30\% of the TNOs could be contact binaries \citep{Sheppard2004, Lacerda2011}. \citet{Sheppard2004} discovered the first contact binary in the trans-Neptunian belt: (139775) 2001~QG$_{298}$. Recently, \citet{Lacerda2014} suggested that the large lightcurve amplitude of the TNO 2003~SQ$_{317}$ could be explained by a contact binary, but they could not totally discard the option of a single, very elongated object. Therefore, 2003~SQ$_{317}$ is a potential contact binary (see \citet{Lacerda2014} for more details). 

In conclusion, to date only one contact binary and one potential contact binary have been found in the trans-Neptunian belt despite their expected high abundance estimate.


\section{Observations and data reduction}
 
We present data obtained with the Lowell Observatory's 4.3~m Discovery Channel Telescope (DCT). Images were obtained using the Large Monolithic Imager (LMI) which is a 6144$\times$6160 CCD \citep{Levine2012}. The total field of view is 12.5$\arcmin$$\times$12.5$\arcmin$ with a pixel scale of 0.12$\arcsec$/pixel (unbinned). Images were obtained using the 3$\times$3 binning mode. Observations were carried out in-situ. 

We always tracked the telescope at sidereal speed. Exposure times of 600 to 700~seconds and the VR-filter (broadband filter to maximize the signal-to-noise ratio of the object) were used. Visual magnitudes\footnote{Visual magnitudes from the Minor Planet Center.} of 2004~TT$_{357}$ were 22.6~mag and 23~mag during our runs in 2015 and 2017, respectively. All relevant geometric information about the observed object at the date of observation, and the number of images are summarized in Table~\ref{Tab:Log_Obs}.

We used the standard data calibration and reduction techniques described in \citet{Thirouin2016, Thirouin2014, Thirouin2012}. The time-series photometry of each target was inspected for periodicities by means of the Lomb periodogram technique \citep{Lomb1976} as implemented in \cite{Press1992}. We also verified our results by using the Phase Dispersion Minimization (PDM, \citet{Stellingwerf1978}).

\begin{table}[h!]
\caption{\label{Tab:Log_Obs} UT-Dates (MM/DD/YYYY), heliocentric (r$_{h}$), and geocentric ($\Delta$) distances in astronomical units (AU) and phase angle ($\alpha$, in degrees) of the observations are reported. We also indicate the number of images (Nb.) obtained each night. The Lowell's Discovery Channel Telescope has been used for all observing runs reported here. Images were obtained with the VR-filter.  }
 
\center
\begin{tabular}{ccccc} 

\hline
UT-date & Nb.   & r$_h$ &  $\Delta$ & $\alpha$   \\
 &       &  [AU]  &  [AU]  &  [$^{\circ}$]     \\
\hline
\hline
12/03/2015  &  32 & 32.243  & 31.260 & 0.1    \\
02/02/2017  &  32 &   32.434 & 32.011-32.015 & 1.6  \\

\hline
\hline

\end{tabular}
 
\end{table}


\section{Photometric results}
\label{sec:photo}

Our dataset is composed of two observing runs, one in 2015 and one in 2017. During our observations in 2015, 2004~TT$_{357}$ was close to opposition and thus its phase angle was 0.1$^\circ$, whereas the 2017 dataset was obtained at higher phase angle, 1.6$^\circ$. Observing one maximum and one minimum over a single night in 2015 allowed us to constrain the rotational period (P$>$7.5~h, assuming a double-peaked lightcurve) and the lightcurve amplitude ($\Delta m$$>$0.7~mag). The 2017 partial lightcurve confirms these estimates. There is no significant change in the lightcurve amplitude between the two datasets. Our photometry is available in Table~\ref{Tab:Summary_photo}. 

\begin{figure*}
\includegraphics[width=9cm, angle=0]{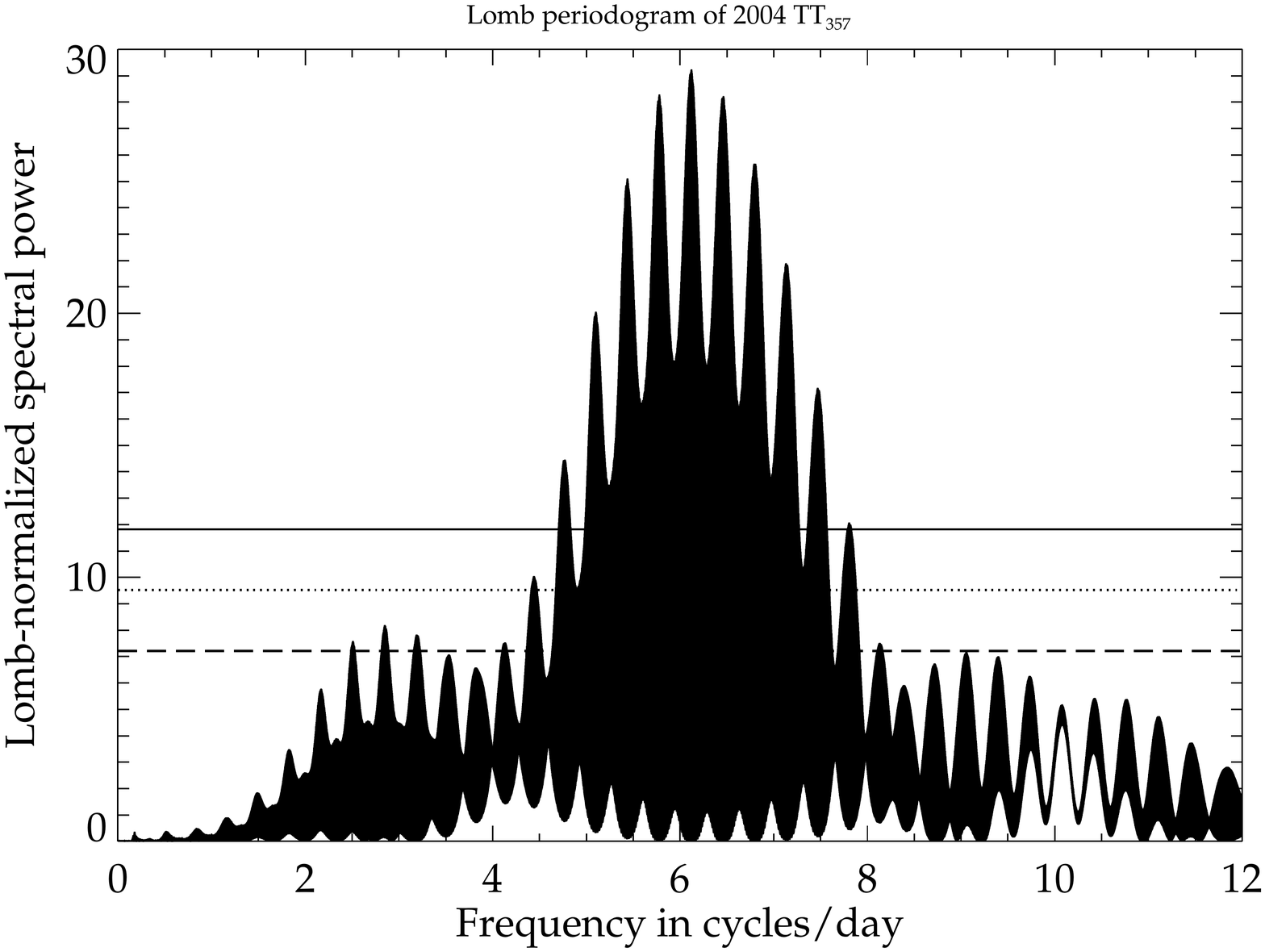}
\includegraphics[width=9cm, angle=0]{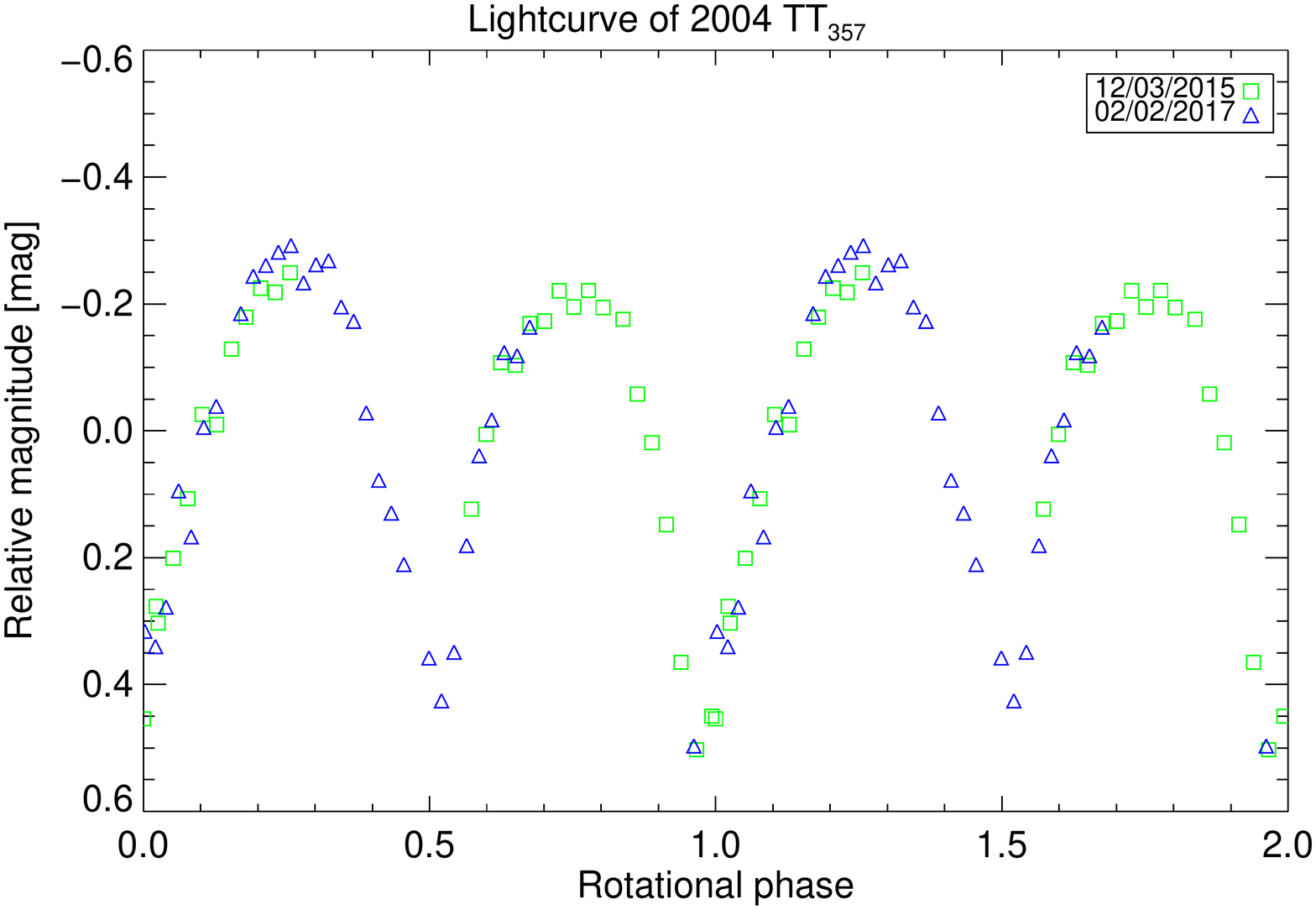}
\includegraphics[width=9cm, angle=0]{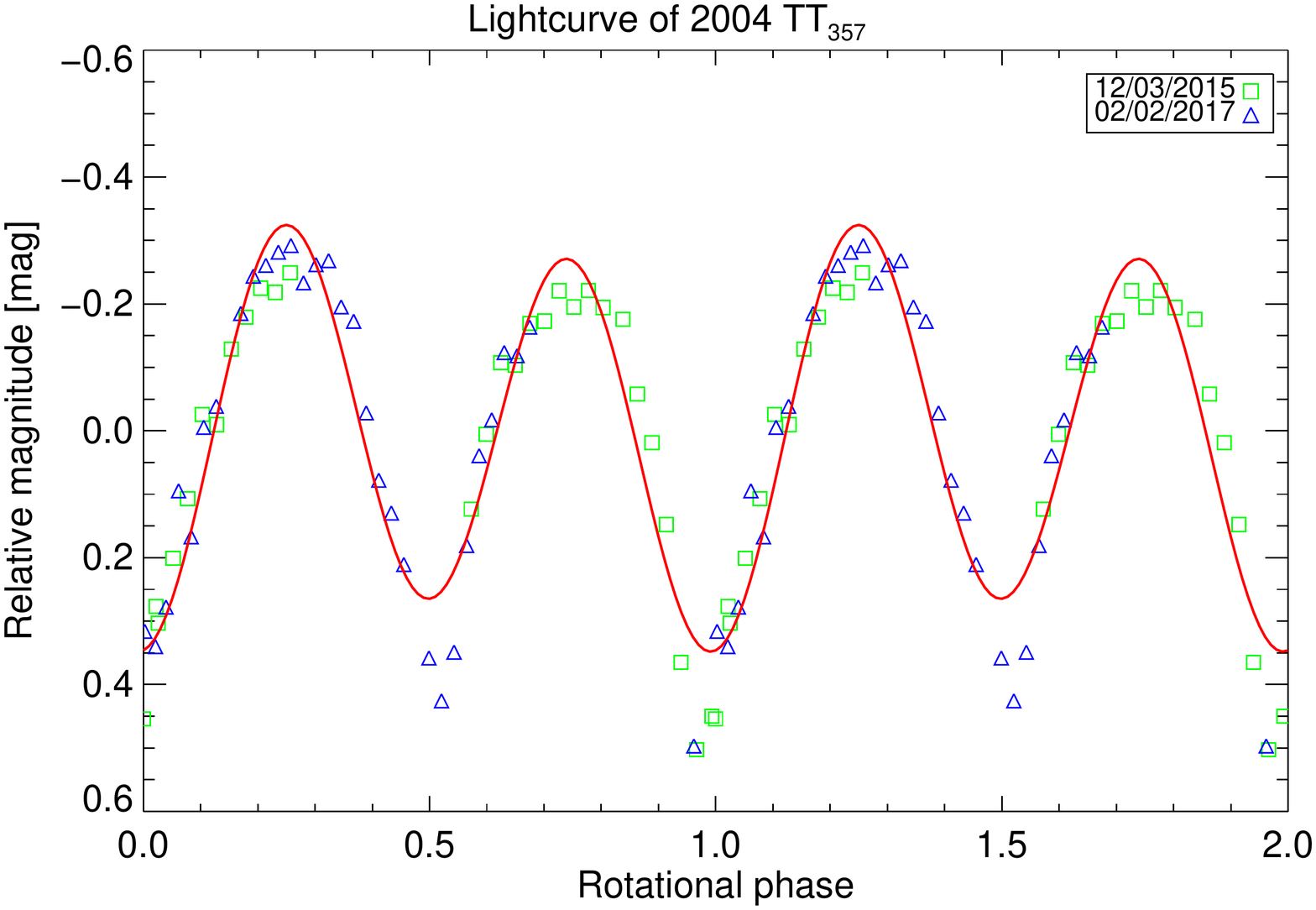}
\caption{\textit{Lomb periodogram and lightcurve of 2004~TT$_{357}$}: Dot, dash and continuous lines are the confidence level at the 90$\%$, 99$\%$, and 99.9$\%$ (respectively). Several peaks are above the 99.9$\%$. The main peak (or peak with the highest spectral power) is located at 3.90~h (6.16~cycles/day). The double-peaked lightcurve (with a rotational period of 7.79~h) is plotted over two cycles (rotational phase between 0 and 2). Error bars of the photometry are not plotted for clarity. We fitted our data with a 2$^{nd}$ order Fourier series, but the fit failed to reproduce the V-and U-shape of the curve (red continuous line).} 
\label{fig:LCLomb}
\end{figure*}

Light-time correction has been applied in order to merge our two observing runs. Our merged dataset has been inspected for periodicity by means of the Lomb periodogram (Figure~\ref{fig:LCLomb}). The Lomb periodogram presents several peaks above the 99.9$\%$ confidence level. The highest peak is located at 6.16~cycles/day (3.89~h), and the main aliases are at 5.78~cycles/day (4.15~h), and 6.48~cycles/day (3.70~h). The PDM method confirms the main peak at 3.89~h. For the rest of our study, we will consider the main peak as the rotational period of the target and will not consider the aliases. Once the period has been identified, we have to choose between the single-peaked lightcurve with a rotational period of 3.89~h and the double-peaked lightcurve with a rotational period of 2$\times$3.89=7.79~h. 

Assuming a triaxial ellipsoid, we have to expect a lightcurve with two maxima and two minima, corresponding to a full rotation of the object. However, if the object is a spheroid with albedo variation(s) on its surface, we have to expect a lightcurve with one maximum and one minimum \citep{Thirouin2014}. A single-peaked lightcurve with a variability of about 0.75~mag would require very strong albedo variation(s) on the object's surface, which is unlikely. On the other hand, assuming a fast rotation of 3.89~h, the object would be deformed due to its rotation and thus its elongated shape would produce a double-peaked lightcurve (as it is the case for Haumea \citep{Lacerda2008, Thirouin2010}). Assuming that 2004~TT$_{357}$ is a strenghtless spherical object with a rotational period of 3.89~h, its density is about 0.7~g~cm$^{-3}$. Finally, by plotting the double-peaked lightcurve (Figure~\ref{fig:LCLomb}), we note that the lightcurve is asymmetric by about 0.05-0.1~mag. For all those reasons, we favor the double-peaked lightcurve for 2004~TT$_{357}$. The lightcurve has a rotational period of 7.79$\pm$0.01~h\footnote{Error bar for the rotational period is the width of the main peak} and a peak-to-peak amplitude of 0.76$\pm$0.03~mag (Figure~\ref{fig:LCLomb}). The lightcurve is plotted over two cycles (i.e. rotational phase between 0 and 2). Error bars are not plotted for clarity.

The lightcurve presents the typical V-/U-shape at the minimum/maximum of brightness characteristic of a contact binary system \citep{Sheppard2004}. However, the lightcurve amplitude is below the 0.9~mag amplitude threshold\footnote{Lightcurve with an amplitude $>$0.9~mag can only be explained by a tidally distorted binary system in hydrostatic equilibrium \citep{Weidenschilling1980, Leone1984}. } needed to infer that this object is a contact binary \citep{Weidenschilling1980, Leone1984}. With a lightcurve amplitude of 0.76~mag, 2004~TT$_{357}$ is between the Roche and Jacobi sequences, and thus its lightcurve can be explained by a very elongated object or by a contact binary assuming that the variability is due to the object's shape (more details in Section~\ref{sec:analysis}). Although a non-equal-sized contact binary may never reach the 0.9~mag amplitude threshold, even when viewed equator-on.


\section{Analysis}
\label{sec:analysis}

The lightcurve of a rotating small body can be produced by: i) albedo variation(s), ii) non-spherical shape, and/or iii) contact/close binary. In this section, we present the arguments in favor of a single very elongated object, a contact binary configuration, and an object with an extreme albedo variation for interpreting the lightcurve of 2004~TT$_{357}$. 

\subsection{Albedo variation(s)}

In the case of asteroids and TNOs, the albedo contributions are usually about 10$\%$-20$\%$ \citep{Degewij1979, Sheppard2004, Sheppard2008, Thirouin2010, Thirouin2014}. In the case of Pluto, the contribution is up to 30$\%$ \citep{Buie1997}. It seems unlikely that 2004~TT$_{357}$ can present an albedo variation of about 80$\%$. Therefore, albedo variations are likely not able to explain the extreme lightcurve amplitude of this object or its asymmetry.
 
\subsection{Elongated shape: Jacobi ellipsoid}

If 2004~TT$_{357}$ is a single elongated object (i.e. Jacobi ellipsoid), its lightcurve amplitude can constrain its elongation, and density.
 
According to \citet{Binzel1989}, if a minor body is a triaxial ellipsoid with axes a$>$b$>$c and rotating along its shortest axis (c-axis), the lightcurve amplitude ($\Delta$${m}$) varies as a function of the observational (or viewing) angle $\xi$ as: 
\begin{equation}
\Delta m = 2.5 \log \left(\frac{a}{b}\right) - 1.25 \log \left(\frac{a^2 \cos ^2 \xi + c^2 \sin ^2 \xi}{b^2 \cos ^2 \xi + c^2 \sin ^2 \xi}\right)
\end{equation}
The lower limit for the object elongation (a/b) is obtained assuming an equatorial view ($\xi$=90$^\circ$).  Considering a viewing angle of $\xi$=90$^\circ$, we estimate an elongation of a/b=2.01, and an axis ratio\footnote{The axis ratio c/a has been derived from \citet{Chandrasekhar1987} work} c/a=0.38. This corresponds\footnote{Assuming that the object is triaxial with semi-axes a$>$b$>$c, and viewed from its equator, the equivalent radius is: \begin{equation} R_{eq}=\sqrt{\frac{ca+cb}{2}} \end{equation}} to a=204~km (a=82~km), b=102~km (b=41~km), and c=78~km (c=31~km) for an albedo of 0.04 (0.25) and an equatorial view.

However, for a random distribution of spin vectors, the probability of viewing an object on the angle range [$\xi$, $\xi$+d$\xi$] is proportional to sin($\xi$)d$\xi$, and the average viewing angle is $\xi$=60$^\circ$ \citep{Sheppard_phd}. Using a viewing angle of 60$^\circ$, we derive an axis ratio a/b$>$2.31. However, ellipsoids with a/b$>$2.31 are unstable to rotational fission \citep{Jeans1919}. Therefore, assuming that 2004~TT$_{357}$ is stable to rotational fission (i.e. a/b$<$2.31), its viewing angle must be larger than 75$^\circ$. 

If 2004~TT$_{357}$ is a triaxial ellipsoid in hydrostatic equilibrium, we can compute its lower density limit based on \citet{Chandrasekhar1987}. Considering an equatorial view, we estimate a density $\rho$$\geq$0.78~g~cm$^{-3}$. This density is typical in the trans-Neptunian belt, and suggests an icy composition for this object \citep{Sheppard2008, Grundy2012, Thirouin2014}. 

We fitted our data with a Fourier series (second-order). This kind of fit is generally used to reproduce lightcurves due to Jacobi ellipsoid \citep{Thirouin2016, Thirouin2014}. But, the fit failed to reproduce the lightcurve, and especially the V-shape of the curve (Figure~\ref{fig:LCLomb}). Therefore, the lightcurve cannot be reproduced if 2004~TT$_{357}$ is assumed to be a Jacobi ellipsoid. One may invoke the fact that strong albedo variations on the object's surface can match the part of the curve that the fit is not able to reproduce \citep{Lacerda2014}. However, such strong variations are unlikely and would have to be located exactly at the maximum and the minimum of brightness of the object.   

\subsection{Contact binary: Roche system}

If 2004~TT$_{357}$ is a contact binary (i.e. Roche system), we can constrain the mass ratio, the separation, the density and the axis ratios of the components. 

\begin{figure*}
\includegraphics[width=12cm, angle=0]{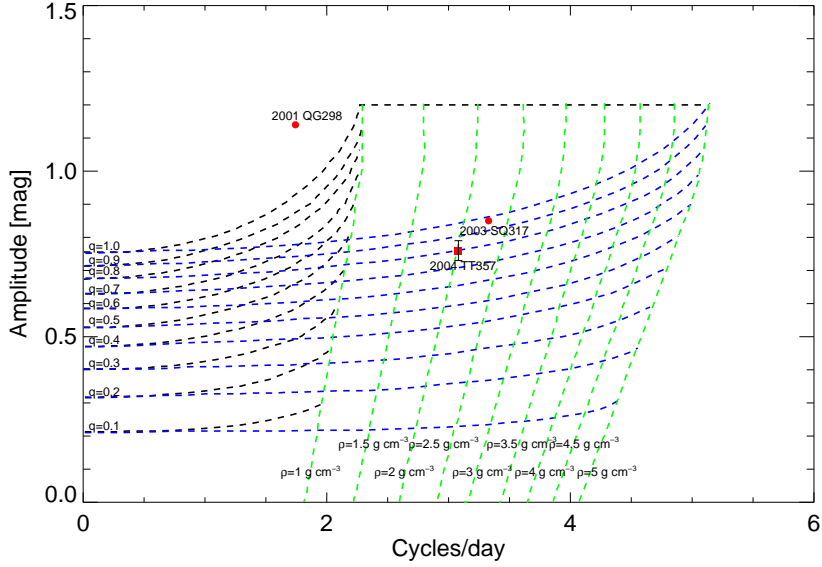}
\includegraphics[width=12cm, angle=0]{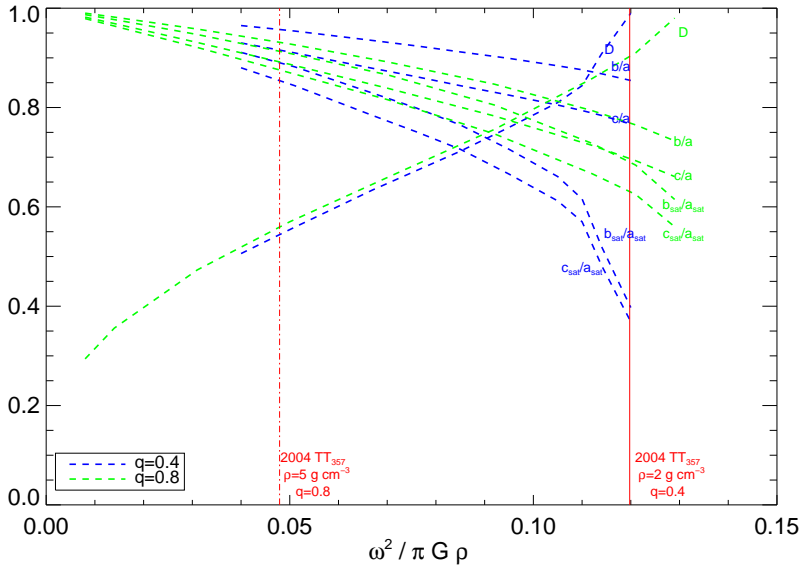}
\caption{\textit{The network of Roche sequences (upper plot), and axis ratios of the components, and parameter D (lower plot)}: For each value of mass ratio (q), black discontinuous lines correspond to a density $\rho$=1~g~cm$^{-3}$, and the blue ones to $\rho$=5~g~cm$^{-3}$. The green lines correspond to the maximum of lightcurve amplitude reached for each density and mass ratio. Figure adapted from Figure 2 in \citet{Leone1984}. Following \citet{Leone1984} formalism, 2004~TT$_{357}$ can have a mass ratio of 0.8, and a density of 5~g~cm$^{-3}$, or a mass ratio of 0.45$\pm$0.05 and a density of 2~g~cm$^{-3}$. We also report 2003~SQ$_{317}$ and 2001~QG$_{298}$ for comparison. Density of 2001~QG$_{298}$ is less than 1~g~cm$^{-3}$, but densities of 2003~SQ$_{317}$ and 2004~TT$_{357}$ are comparable. Axis ratios of the primary (b/a, c/a), of the secondary (b$_{sat}$/a$_{sat}$, c$_{sat}$/a$_{sat}$), and parameter (D) for a mass ratio of 0.4 (blue), and 0.8 (green) are also plotted.  The red dot-dash line is 2004~TT$_{357}$ assuming a mass ratio of 0.8, and the red continuous line is 2004~TT$_{357}$ assuming a mass ratio of 0.4.  }
\label{fig:Roche}
\end{figure*}

\citet{Leone1984} studied the sequences of equilibrium of these binaries with a mass ratio between 0.01 and 1 (see \citet{Leone1984} for more details about the model\footnote{Main hypothesis: i) object is seen equator-on, and ii) phase angle is 0$^\circ$. Similar criteria are assumed in \citet{Lacerda2014}.}). Using the network of Roche sequences in the plane \textit{lightcurve amplitude-rotational frequency}, we can estimate the mass ratio and the density of 2004~TT$_{357}$ (Figure~\ref{fig:Roche}, adapted from Figure 2 of \citet{Leone1984}). We derive two main options: i) a system with a mass ratio of q$_{max}$=0.8$\pm$0.05 and a density of $\rho$$_{max}$=5~g~cm$^{-3}$, or ii) a system with a mass ratio of q$_{min}$=0.45$\pm$0.05 and a density of $\rho$$_{min}$=2~g~cm$^{-3}$. We derive q$_{min}$ and q$_{max}$ by taking into account the error bar of the lightcurve amplitude. Based on the fact that we only have one lightcurve of this object and the number of assumptions made by \citet{Leone1984}, we will use conservative mass ratios of q$_{min}$=0.4, and q$_{max}$=0.8. Using \citet{Leone1984}, we are only able to derive the extreme cases (min and max), combination of values in between are also possible. Only a careful modeling of the system using several lightcurves at different epochs will allow us to improve the mass ratio, density as well as geometry of the system.  
The parameter\footnote{$\omega$ is the orbital angular velocity, $\omega$=2$\pi$/P where P is the rotational period. $\rho$ is the density and G is the gravitational constant.} $\omega^{2}$/($\pi$G$\rho$) is 0.048 with a mass ratio of 0.8, and is 0.120 with a mass ratio of 0.4 (Figure~\ref{fig:Roche}). Using the Table 1 of \citet{Leone1984}, we derive the axis ratios and the separation between the components.

If 2004~TT$_{357}$ is a binary system with a mass ratio of 0.8, and a density of 5~g~cm$^{-3}$, we derive the axis ratios of the primary: b/a=0.93, c/a=0.89, the axis ratios of the secondary: b$_{sat}$/a$_{sat}$=0.91, c$_{sat}$/a$_{sat}$=0.88. The parameter D is defined as (a+a$_{sat}$)/d where d is the orbital separation, and a, a$_{sat}$ are the longest axes of the primary and of the secondary, respectively. The components are in contact when D=1. We calculate that D=0.56, thus d=(a+a$_{sat}$)/0.56.

If 2004~TT$_{357}$ is a binary system with a mass ratio of 0.4, and a density of 2~g~cm$^{-3}$, we derive the axis ratios of the primary: b/a=0.85, c/a=0.77 (a=75/30~km, b=63/25~km, and c=57/22~km assuming an albedo of 0.04/0.25), the axis ratios of the secondary: b$_{sat}$/a$_{sat}$=0.40, c$_{sat}$/a$_{sat}$=0.37 (a=89/37~km, b=75/31~km, and c=68/28~km assuming an albedo of 0.04/0.25). The parameter D is 1, suggesting that the components are in contact.
 
The largest TNOs have higher densities than the smaller ones \citep{Sheppard2008, Grundy2012, Thirouin2013, Brown2013, Vilenius2014}. Considering only the binary/multiple systems with true densities derived from their mutual orbits, the largest objects have a mean density around 2~g~cm$^{-3}$, the intermediate-sized objects have a mean density of about 1.5~g~cm$^{-3}$, whereas the smallest ones have a mean density of $\sim$0.5~g~cm$^{-3}$ \citep{Thirouin2013}. Therefore, a density of 5~g~cm$^{-3}$ is unlikely in the trans-Neptunian belt, and especially for as small an object as 2004~TT$_{357}$. Therefore, we consider that explaining the lightcurve from a contact binary system with a mass ratio of 0.8 and a density of 5~g~cm$^{-3}$ is unrealistic. Even if a density of 2~g~cm$^{-3}$ seems more reasonable for a TNO, it is important to point out that such a density will make 2004~TT$_{357}$ one of the densest objects in its size range. Some TNOs have a density of around 2~g~cm$^{-3}$ and thus such a high value is not uncommon \citep{Thirouin2016}. The densities of 2004~TT$_{354}$, and 2003~SQ$_{357}$ are comparable, whereas the density of 2001~QG$_{298}$ is less than 1~g~cm$^{-3}$. This high density suggests a rocky composition.  \\

In conclusion, we have presented arguments in favor of a very elongated single object or a contact binary system to explain the extreme variability of the lightcurve of 2004~TT$_{357}$. Based on the U-/V-shape morphology of the lightcurve the contact binary explanation seems most likely. More data at different observing angles in the future will allow us to confirm the nature of this object/system \citep{Lacerda2011}.  

\section{Search for companion(s)}

\begin{figure*}
\includegraphics[width=13cm, angle=0]{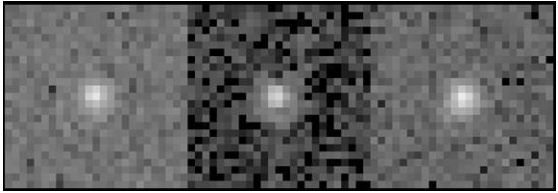}
\caption{\textit{\textit{Hubble Space Telescope} images}: Three 25$\times$25~pixel image sections centered on 2004~TT$_{357}$ are shown in the order they were exposed. Filters and exposures used were F606W/300~s, F814W/400~s, and F606W/300~s respectively from left to right. All are shown with the same log stretch. There is no visual or PSF evidence of a resolved or unresolved binary. }
\label{fig:HST}
\end{figure*}

2004~TT$_{357}$ was observed with the \textit{Hubble Space Telescope} in order to search for a binary companion. Images were obtained on 21 February 2012, starting at 09:46~UT in visit 86 of cycle 19 GO snapshot program 12468, \textit{How Fast Did Neptune Migrate? A Search for Cold Red Resonant Binaries}. Images were obtained using WFC3 with two 350~s exposures in the F606W filter and one 400~s exposure in the F814W. The individual flat-fielded images show a single, compact point spread function (PSF) with a uniform appearance in all three images. An empirical PSF-fit yields an average full width at half maximum (FWHM) of 1.70~pixels, consistent with an unresolved, single object (Figure~\ref{fig:HST}). 


\section{Summary and Conclusions}
 
We have collected photometric data for 2004~TT$_{357}$ using the Lowell's Discovery Channel Telescope in 2015 and 2017. Our results are summarized here: 

\begin{itemize}

\item The lightcurve of 2004~TT$_{357}$ presents a V-/U-shape at the minimum/maximum of brightness. 2004~TT$_{357}$ has an asymmetric double-peaked lightcurve with a rotational period of 7.79~h, and an amplitude of 0.76~mag. Such a large lightcurve amplitude can be explained by a very elongated single object or a contact/close binary.  

\item In the case of a contact binary, we find two extreme solutions: i) a system with a mass ratio q$_{min}$=0.45$\pm$0.05, a density $\rho$$_{min}$=2~g cm$^{-3}$ or ii) a system with a mass ratio q$_{max}$=0.8$\pm$0.05, a density $\rho$$_{max}$=5~g cm$^{-3}$. Because a density of $\rho$=5~g cm$^{-3}$ is unlikely in the trans-Neptunian belt for an object in the size range of 2004~TT$_{357}$, we favor the solution given by a mass ratio of about 0.4, and a density of $\rho$=2~g cm$^{-3}$. This is still a higher than normal density for a small TNO. 

\item Assuming a mass ratio of 0.4, and a density of $\rho$=2~g cm$^{-3}$, we estimate the axis ratio of the primary as b/a=0.85, c/a=0.77, and b$_{sat}$/a$_{sat}$=0.40, c$_{sat}$/a$_{sat}$=0.37 for the secondary. We derive a parameter D=1 suggesting that the components are in contact.

\item If 2004~TT$_{357}$ is a Jacobi ellipsoid in hydrostatic equilibrium, its density is $\rho$$\geq$0.78~g cm$^{-3}$, and its elongation is a/b=2.01 (assuming a viewing angle $\xi$=90$^\circ$). In order for this object to be rotationally stable, the viewing angle needs to be between 75$^\circ$ and 90$^\circ$. 

\item Only changes in the lightcurve in the future can allow us to further confirm the binary nature (or not) of 2004~TT$_{357}$. Lightcurve(s) at different epoch(s) will be needed to model and characterize the system and its geometry (similar work published in \citet{Lacerda2011} for 2001~QG$_{298}$). 

\item No resolved companion orbiting 2004~TT$_{357}$ has been found based on \textit{Hubble Space Telescope} data obtained in 2012.  

\end{itemize}

%
 
\acknowledgments

We thank the referee for her/his useful comments. This research is based on data obtained at the Lowell Observatory's Discovery Channel Telescope (DCT). Lowell operates the DCT in partnership with Boston University, Northern Arizona University, the University of Maryland, and the University of Toledo. Partial support of the DCT was provided by Discovery Communications. LMI was built by Lowell Observatory using funds from the National Science Foundation (AST-1005313). We acknowledge the DCT operators: Heidi Larson, Teznie Pugh, and Jason Sanborn. Audrey Thirouin is partly supported by Lowell Observatory funding.\\

 \clearpage

\clearpage

\begin{deluxetable}{lccc}
 
\tablecaption{\label{Tab:Summary_photo} Photometry used in this paper is available in the following table. Julian date is without light-time correction. }
\tablewidth{0pt}
\tablehead{
Object  & Julian Date & Relative magnitude &  Error  \\
        &          &   [mag]           &  [mag]\\
}
\startdata  
2004~TT$_{357}$ &   &   &   \\
 & 2457359.63593 & 0.45 & 0.03  \\
 & 2457359.64421 & 0.30 & 0.04  \\ 
& 2457359.65277 & 0.20 & 0.04  \\ 
& 2457359.66103 & 0.11 & 0.04  \\ 
& 2457359.66931 & -0.03 & 0.03  \\ 
& 2457359.67757 & -0.01 & 0.03  \\ 
& 2457359.68584 & -0.13 & 0.03  \\ 
& 2457359.69411 & -0.18 & 0.03  \\ 
& 2457359.70236 & -0.22 & 0.03  \\ 
& 2457359.71062 & -0.22 & 0.03  \\ 
& 2457359.71917 & -0.25 & 0.03  \\ 
& 2457359.82174 & 0.12 & 0.04  \\ 
& 2457359.83029 & 0.01 & 0.04  \\ 
& 2457359.83854 & -0.11 & 0.04  \\ 
& 2457359.84683 & -0.10 & 0.04  \\
 & 2457359.85509 & -0.17 & 0.04  \\ 
& 2457359.86337 & -0.17 & 0.04  \\ 
& 2457359.87163 & -0.22 & 0.05  \\ 
& 2457359.87988 & -0.20 & 0.06  \\ 
& 2457359.88815 & -0.22 & 0.06  \\ 
& 2457359.89640 & -0.19 & 0.07  \\ 
& 2457359.90770 & -0.18 & 0.10  \\ 
& 2457359.91596 & -0.06 & 0.05  \\
 & 2457359.92422 & 0.02 & 0.04  \\ 
& 2457359.93248 & 0.15 & 0.05  \\ 
& 2457359.94074 & 0.36 & 0.08  \\ 
& 2457359.94947 & 0.50 & 0.11  \\ 
& 2457359.95808 & 0.45 & 0.10  \\ 
& 2457359.96749 & 0.28 & 0.10  \\ 
& 2457786.59029 & 0.50 & 0.03  \\ 
& 2457786.60347 & 0.32 & 0.05  \\ 
& 2457786.60950 & 0.34 & 0.04  \\ 
& 2457786.61550 & 0.28 & 0.05  \\ 
& 2457786.62262 & 0.09 & 0.05  \\
 & 2457786.62975 & 0.17 & 0.04  \\ 
& 2457786.63688 & -0.01 & 0.05  \\ 
& 2457786.64399 & -0.04 & 0.04  \\ 
& 2457786.65787 & -0.19 & 0.01  \\ 
& 2457786.66497 & -0.24 & 0.04  \\ 
& 2457786.67209 & -0.26 & 0.02  \\ 
& 2457786.67919 & -0.28 & 0.02  \\ 
& 2457786.68632 & -0.29 & 0.02  \\
 & 2457786.69343 & -0.23 & 0.02  \\ 
& 2457786.70054 & -0.26 & 0.04  \\ 
& 2457786.70765 & -0.27 & 0.03  \\ 
& 2457786.71477 & -0.20 & 0.04  \\ 
& 2457786.72187 & -0.17 & 0.03  \\ 
& 2457786.72900 & -0.03 & 0.05  \\ 
& 2457786.73610 & 0.08 & 0.03  \\ 
& 2457786.74323 & 0.13 & 0.03  \\
 & 2457786.75032 & 0.21 & 0.05  \\ 
& 2457786.76456 & 0.36 & 0.04  \\ 
& 2457786.77168 & 0.43 & 0.04  \\ 
& 2457786.77878 & 0.35 & 0.05  \\ 
& 2457786.78591 & 0.18 & 0.04  \\
 & 2457786.79301 & 0.04 & 0.05  \\
 & 2457786.80014 & -0.02 & 0.05  \\ 
& 2457786.80723 & -0.12 & 0.04  \\ 
& 2457786.81456 & -0.12 & 0.04  \\ 
& 2457786.82168 & -0.16 & 0.05  \\ 
\enddata

\end{deluxetable}

\end{document}